\documentclass[twocolumn,showpacs,preprintnumbers,amsmath,amssymb]{revtex4}

\usepackage{graphicx}
\usepackage{dcolumn}
\usepackage{bm}

\begin{document}

\preprint{APS/123-QED}

\title{Magnetic tunneling junctions with the Heusler compound Co$_2$Cr$_{0.6}$Fe$_{0.4}$Al}

\author{A. Conca}
 \email{conca@uni-mainz.de}
 \homepage{http://www.uni-mainz.de/FB/Physik/AG_Adrian/}
\author{S. Falk}
\author{G. Jakob}
\author{M. Jourdan}
\author{H. Adrian}
\affiliation{Institut f\"ur Physik, Johannes Gutenberg-Universit\"at, Staudingerweg 7, 55128 Mainz, Germany}

\date{\today}

\begin{abstract}
Certain Heusler phases belong to the materials which are discussed as potential half metals. Here results of tunneling experiments with the full-Heusler alloy CoCr$_{0.6}$Fe$_{0.4}$Al are presented. 
\newline
The Heusler alloy is used as an electrode  of magnetic tunneling junctions. The junctions are deposited by magnetron dc sputtering using shadow mask techniques with AlO$_{x}$ as a barrier and cobalt as counter electrode.     Measurements of the magnetoresistive differential conductivity in  a temperature range between 4K and 300K  are shown. An analysis of the barrier properties applying the Simmons model to the bias dependent junction conductivity is performed. VSM measurements were carried out to examine the magnetic properties of the samples.
\end{abstract}

\maketitle

\section{Introduction}
Materials with large spin polarization are required for applications in the field of spinelectronics. In this sense, research on compounds which are expected to be half-metals, i.e. with 100 \% spin polarization, is of special interest. Half metals (HMs) are expected to have a gap at E$_{F}$ for one spin band while the other spin band is metallic leading to a completely polarized current. The use of HM electrodes in spin valves or in magnetic tunneling junctions (MTJs) will result in a pronounced increase of the  magnetoresistance. The full Heusler alloy Co$_2$Cr$_{0.6}$Fe$_{0.4}$Al is expected to be a HM. It is used as an electrode in this work, further information on this compound can be found in \cite{gerhard}.
Following the Julli\`ere model \cite{Julliere}, the tunneling magnetoresistance (TMR) ratio of a junction is related to the  spin  polarization $P$ of the electrodes:
\begin{equation}
\label{tmr}
TMR={2P_1P_2 \over {1+P_1P_2}}
\end{equation}
Please note that in this definition the TMR ratio is normalized to the value of resistance in the antiparallel configuration.
The Julli\`ere model  is not able to explain certain features of MTJs like the temperature and bias voltage dependence of the tunneling magnetoresistance. In expression (\ref{tmr}), $P_{n}$ stands for the polarization of the bulk material, but the tunneling conductivity is only sensitive to the local  density of states in the vicinity of the barrier which may differ from the one of the bulk material. Additionally, the Julli\`ere model does not take into account the various direct effects of the barrier on the tunneling probability  \cite{deteresa,mig,ding}. However, the model is commonly used to estimate the spin polarization of the electrodes.
First attempts of building MTJs with half metallic electrodes were discouraging due to the low TMR obtained compared to the theoretically expected value for $\approx$ 100\% spin polarization \cite{tanaka,tanaka2}. Recently, a TMR ratio of 16 \% at room temperature and 26.5 \% at 4K with Co$_2$Cr$_{0.6}$Fe$_{0.4}$Al as an electrode was reported \cite{inomata}.
\begin{figure}[htb]
\includegraphics[width=0.9\linewidth]{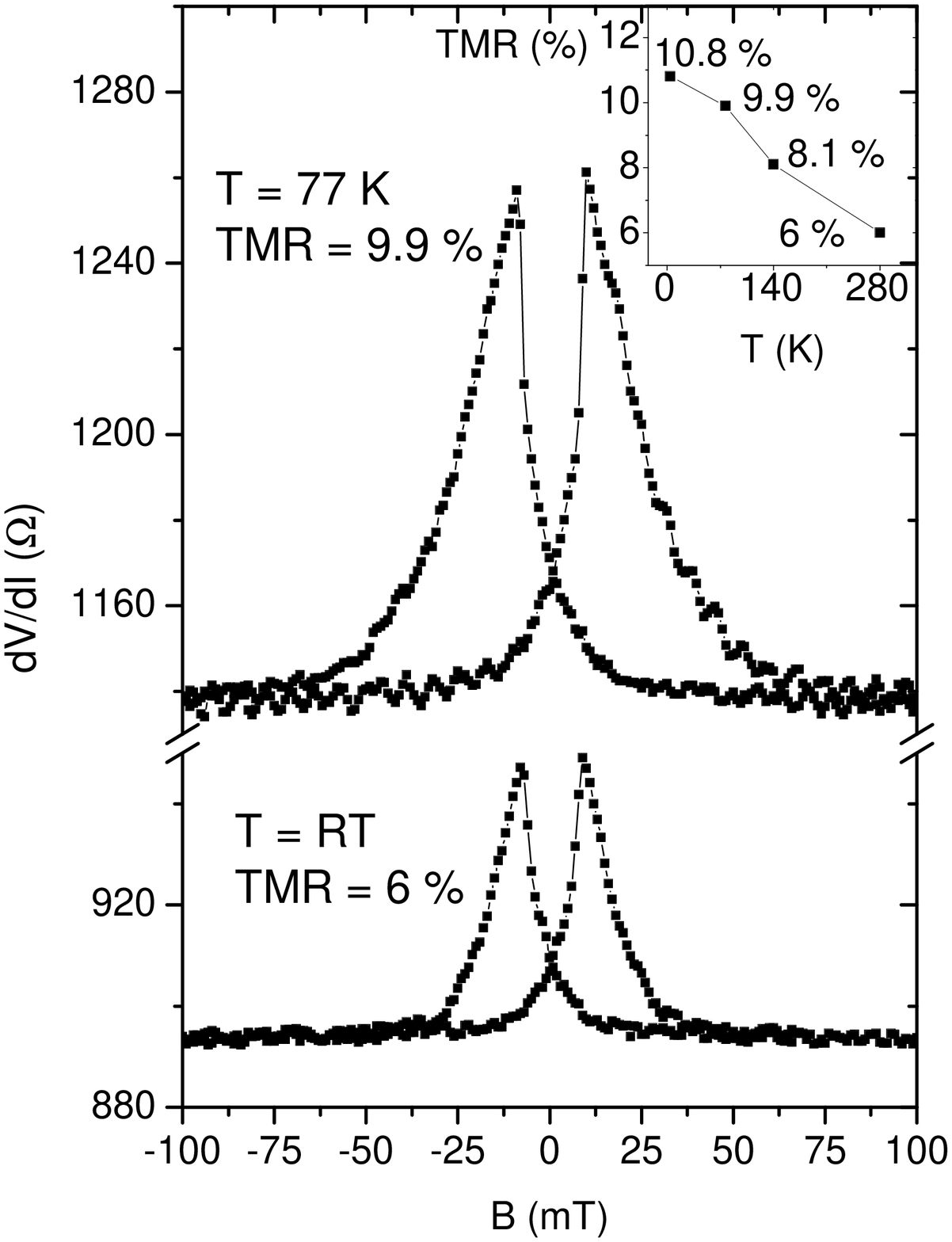}
\caption{\label{mr}\footnotesize Differential resistivity of a Co$_2$Cr$_{0.6}$Fe$_{0.4}$Al/ AlO$_x$/Co junction  at room temperature and 77K. The inset shows the dependence of the TMR ratio on the temperature. A bias voltage of 1 mV was applied.}
\end{figure}
  
\section{Deposition}
The junctions were deposited by magnetron dc sputtering in a chamber with a base pressure of $\sim$10$^{-7}$mbar on Al$_2$O$_3$ (1120) substrates. They were deposited in a cross geometry using metal shadow masks with a width of 150 and 300 $\mu$m. The structure of the layers of the junction is CCFA/AlO$_x$/Co where CCFA stands for Co$_2$Cr$_{0.6}$Fe$_{0.4}$Al. No pinning system was used, the antiparallel configuration is achieved due to the different coercive fields of the electrodes. The film thickness was determined by x-ray reflectometry. The minimum thickness of the CCFA electrode was chosen to be  125nm in order to reduce the resistance of the bottom electrode. It must be emphasized that, due to the high specific resistivity of the Heusler compound ($\rho$ = 230$\mu\Omega\cdot$cm \cite{gerhard}), thick layers are needed to discard geometrical enhancement of the TMR ratio \cite{veerdonk,moodera}. The aluminum layer was oxidized using plasma oxidation at 0.2mbar with oxidation times between 120s and 180s. Oxidation times below 120s result in low junction resistances (below 175$\Omega)$ and it is not possible to discard current distribution effects. The Co layer was deposited with a thickness of 200nm.
CCFA films with best crystallographic properties are grown at 600$^0$C. However, the first attempts of depositing MTJs were done at room temperature to obtain smooth surfaces  to grow a good insulating barrier.
Increased deposition temperatures have not shown large changes in the TMR signal up to 400$^0$C. The preparation of MTJs at temperatures over 400$^0$C resulted in very low or inexistent TMR signal. Possible reasons are an increased roughness of the surface of the Heusler film  and  an enhanced degradation of the surface due to oxygen impurities in the sputtering gas.
\begin{figure}[htb]
\includegraphics[width=0.9\linewidth]{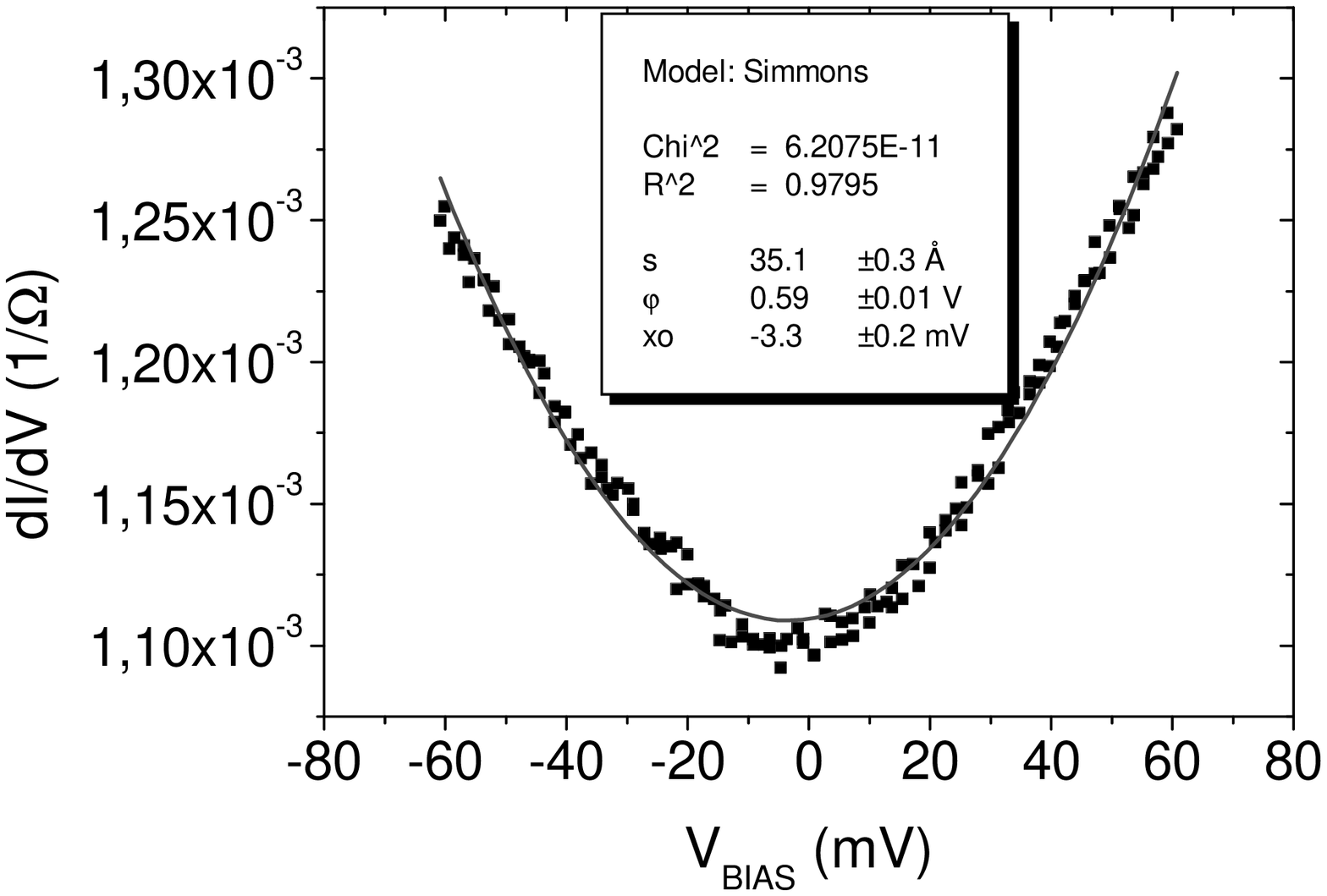}
\caption{\label{simmons}\footnotesize Dependence of the differential conductivity on the bias voltage. The parabolic behavior is a characteristic of tunneling. The solid line represents the fit to the Simmons model. The averaged height ($\phi$) and the effective width (s) of the barrier are given in volts and \aa ngstr\"oms, respectively. x$_0$ is the offset of the parabola from zero bias.}
\end{figure}
  
\section{Tunneling conductivity}
The maximum TMR ratio observed was 10.8 \% at 4K and 6 \% at room temperature. The magnetoresistive curves for 77K and room temperature are shown in fig.\,\ref{mr}. The Heusler electrode was deposited at 300$^0$C and the oxidation time was 180s. The sharpness of the peaks is an indication for the similarity of the coercive fields of the two electrodes. It is very probable that a full antiparallel configuration is not achieved resulting in a reduced TMR amplitude. 
In order to prevent current distribution effects the quotient between the resistance of the junction ($R_J$) and the resistance of the electrodes over the junction area ($R_L$) must be much larger than 1.  This was always the case for our junctions. In the case of the junction shown in fig.\,\ref{mr} the value is $R_J/R_L > 40$.
Previous results of tunneling with the CCFA compound show a larger TMR effect \cite{inomata} with a thickness for the CCFA of 10nm and a junction resistance of 140$\Omega$. From our experience with the high specific resistivity of the Heusler films these values might result in a geometrical enhancement of the TMR ratio.
\begin{figure}[htb]
\includegraphics[width=0.9\linewidth]{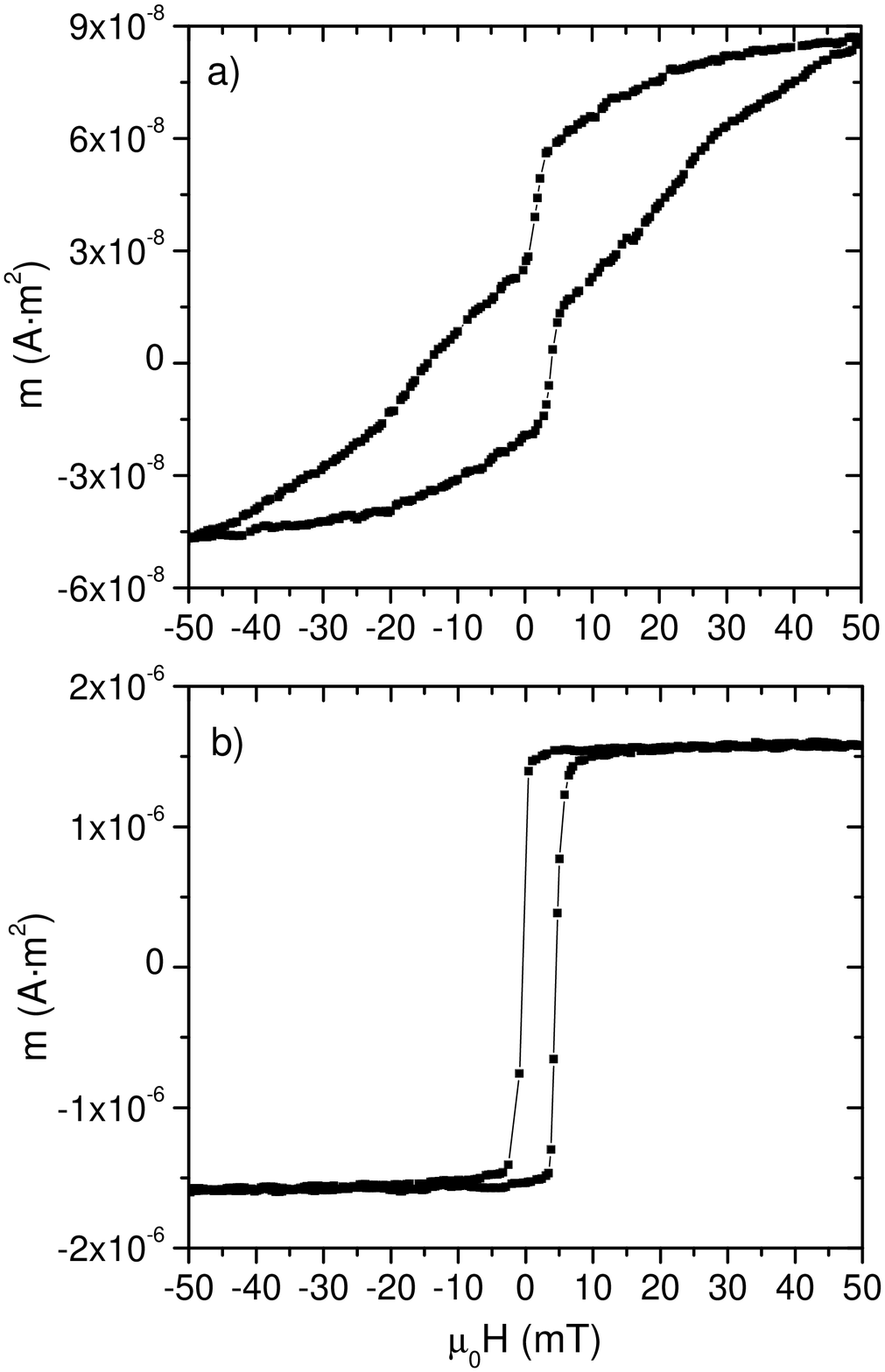}
\caption{\label{vsm}\footnotesize VSM measurements of the magnetization of the samples. The graphs show the hysteresis cycles for a junction deposited with shadows masks (a) and a Heusler film deposited on a 5x5  mm$^2$ substrate covered with an oxidized Al layer (b).}
\end{figure}
A study of the dependence of the differential conductivity dI/dV on the bias voltage was performed for every junction to assure that tunneling was the main conduction channel. The parabolic behavior shown in fig.\,\ref{simmons} for the same junction which provided the results of fig.\,\ref{mr} is an indication of tunneling.  From a fit  using the Simmons model \cite{sim} we can extract the characteristic parameters of  this junction. The effective width of the barrier was s = 35\AA \, with an average height $\varphi$ = 0.6V.  In all measurements it is observed that the parabola is not  centered at zero bias but  has a small offset between 3 and 4 mV. This offset may result from an  asymmetry of the barrier.
  
\section{VSM measurements}  
Several samples were analyzed in a Vibrating Sample Magnetometer (VSM). The hysteresis cycle for the junction of fig.\,\ref{mr} is shown in fig.\,\ref{vsm}a. The current leads were removed by cutting the substrate with a diamond saw. A small part of the leads could not be removed, this part had a length smaller than 1 mm. The offset in the magnetization axis may be a result of an electronic drift due to the low signal of the junction. The reason for the low signal is the small area of the sample. In the graph it is not possible to see a completely independent switching of the magnetization of the layers, in  correspondence with the sharpness of the peaks in the magnetoresistivity measurements. 
A Heusler film with the same thickness as the junction was deposited on a 5x5 mm$^2$ substrate. The film was covered with an Al layer and then oxidized to create the same conditions as in the deposition of a junction. The hysteresis cycle for this sample is shown in fig.\,\ref{vsm}b. By comparison of the two plots it is deduced that the  layer of CCFA switches sharply at relatively small magnetic fields. The switching of the Co layer is broader and at higher magnetic fields.\\
The financial support by the Materials  Science Research Center (MWFZ) Mainz is acknowledged.

\end{document}